\documentclass{webofc}

\usepackage[varg]{txfonts}   
\usepackage{hyperref}
\usepackage{url}
\usepackage{subcaption}
\usepackage[normalem]{ulem} 
\usepackage{feynmp-auto}
\usepackage{tikz}
\usepackage{tikz-feynman}
\usepackage{pgfplots}
\pgfplotsset{compat=1.18}
\usetikzlibrary{patterns,arrows.meta,decorations.pathreplacing}

\hypersetup{colorlinks=true,citecolor=blue,urlcolor=blue,linkcolor=blue}
\newcommand\kala{\ensuremath{\kappa_{\lambda}}}

\def\reffi#1{\mbox{Fig.~\ref{#1}}}

\newcommand\citere[1]{Ref.~\cite{#1}}
\newcommand\citeres[1]{Refs.~\cite{#1}}

\begin{document}
\begin{flushright}
\texttt{DESY-26-025}\\
\texttt{IFT–UAM/CSIC-26-017}\\
\texttt{FR-PHENO-2026-005}
\end{flushright}


%
\title{Sensitivity to new physics: single-Higgs couplings vs.\ the trilinear Higgs coupling}
%
%

\author{
        \firstname{Henning}
        \lastname{Bahl}\inst{1}\fnsep\thanks{\email{bahl@thphys.uni-heidelberg.de}} \and
        \firstname{Johannes} \lastname{Braathen}\inst{2}\fnsep\thanks{\email{johannes.braathen@desy.de}} \and
        \firstname{Martin} \lastname{Gabelmann}\inst{3}\fnsep\thanks{\email{martin.gabelmann@physik.uni-freiburg.de}} \and
        \firstname{Sven} \lastname{Heinemeyer}\inst{4}\fnsep\thanks{\email{sven.heinemeyer@cern.ch}} \and
        \firstname{Kateryna} \lastname{Radchenko Serdula}\inst{4}\fnsep\thanks{\email{kateryna.radchenko@desy.de}} \and
        \firstname{Alain} \lastname{Verduras Schaeidt}\inst{2}\fnsep\thanks{\email{alain.verduras@desy.de}}\and
        \firstname{Georg} \lastname{Weiglein}\inst{2,5}\fnsep\thanks{\email{georg.weiglein@desy.de}}}
\institute{
Institut für Theoretische Physik, Universit{\"a}t Heidelberg, Philosophenweg 16, 69120 Heidelberg, Germany
\and
Deutsches Elektronen-Synchrotron DESY, Notkestr.~85,  22607  Hamburg,  Germany
\and 
Albert-Ludwigs-Universit{\"a}t Freiburg, Physikalisches Institut, Hermann-Herder-Str.~3, 79104 Freiburg, Germany
\and
Instituto de F\'isica Te\'orica (UAM/CSIC), Universidad Aut\'onoma de Madrid, Cantoblanco 28049 Madrid, Spain
\and
Institut für Theoretische Physik, Universität Hamburg, Luruper Chaussee 149, 22761 Hamburg, Germany
}

\abstract{
The trilinear Higgs self-coupling provides a unique probe of the structure of the Higgs potential and of the nature of the electroweak phase transition, and constitutes a key target for future collider experiments. Recent studies have shown that confronting theoretical predictions for the trilinear Higgs coupling with current experimental bounds offers a powerful and complementary way to test effects of physics beyond the Standard Model (BSM), in particular those arising from extended Higgs sectors. Meanwhile, substantial progress has been achieved in the precise calculation and automation of the trilinear Higgs coupling in a wide class of BSM models.
This contribution discusses several BSM scenarios, compatible with existing constraints, in which sizeable deviations in the trilinear Higgs coupling 
w.r.t.\ the Standard Model (SM) value are predicted, while other Higgs observables remain close to their SM expectations and are therefore difficult to probe experimentally. 
These results highlight the strong physics motivation for a precise measurement of the trilinear Higgs coupling at a future Higgs factory.
}
\maketitle

\pagestyle{plain}

\noindent\textit{Talk presented at the International Workshop on Future Linear Colliders 2025 (LCWS2025), Valencia, Spain}
\newpage

\section{Introduction}
\setcounter{page}{1}
\label{intro}

The properties of the Higgs boson discovered in 2012, with a measured mass of about 125~GeV, provide strong evidence for the mechanism responsible for mass generation. Measurements of the couplings of the Higgs boson to SM particles are in agreement with the Standard Model (SM) predictions at the 
level of $\sim 10\%$. However, a more complete determination of the Higgs-boson sector
requires direct information on the shape of the scalar potential, of which only the position of the minimum and the local curvature around it are currently known. A crucial next step in probing the scalar potential is the measurement of the trilinear Higgs coupling $\lambda_{hhh}$, or equivalently its coupling modifier $\kappa_{\lambda} \equiv \lambda_{hhh}/\lambda_{hhh}^{\mathrm{SM},\,(0)}$ (where $\lambda_{hhh}^{\mathrm{SM},\,(0)}$ is the tree-level prediction in the SM). At present, it is constrained to be in the range $-0.71 < \kappa_{\lambda} < 6.1$ by the combination of ATLAS and CMS Run~2 searches for Higgs-boson pair production 
~\cite{CMS:2026nuu}.

The shape of the scalar potential does not only shed light on the number of scalar degrees of freedom present in our Universe, but may also facilitate an explanation of the observed matter–antimatter asymmetry. Depending on the nature of the electroweak symmetry–breaking transition, the baryon asymmetry of the Universe can be generated through the scenario of electroweak baryogenesis. In this hypothetical framework, the transition from a metastable (false) vacuum to the electroweak vacuum proceeds via a first-order phase transition, which, moreover, must be sufficiently strong. The strength of the transition is commonly quantified by the parameter $\xi_n = v_n(T_n)/T_n \gtrsim 1$, where $v_n$ is the value of the vacuum expectation value at the nucleation temperature $T_n$. A strong first-order electroweak phase transition (SFOEWPT) is a necessary condition for successful baryogenesis and is typically (for single-step transitions) associated with deviations of the trilinear Higgs coupling from its SM value~\cite{Kanemura:2004ch,Grojean:2004xa}. For instance, in the Two Higgs Doublet Model (2HDM), the realisation of a SFOEWPT requires values of $\kappa_{\lambda}\sim 2$~\cite{Biekotter:2022kgf}. These findings are not restricted to the 2HDM: e.g. in the general singlet extension of the SM, the occurrence of SFOEWPTs driven by the SM-like doublet field imply values of $\kappa_\lambda$ around 1.5~\cite{Braathen:2025svl}.

Such large deviations of the trilinear coupling from the SM prediction can be realised in particular as a consequence of the presence of heavy new scalar states that contribute to higher-order radiative corrections~\cite{Kanemura:2002vm,Kanemura:2004mg,Aoki:2012jj,Kanemura:2015fra,Kanemura:2015mxa,Hashino:2015nxa,Kanemura:2016lkz,Kanemura:2017wtm,Kanemura:2017gbi,Chiang:2018xpl,Senaha:2018xek,Braathen:2019pxr,Kanemura:2019slf,Braathen:2019zoh,Braathen:2020vwo,Bahl:2022jnx,Bahl:2022gqg,Bahl:2023eau,Aiko:2023nqj,Basler:2024aaf,Bahl:2025wzj,Braathen:2025qxf}. Even in scenarios where the properties of the Higgs boson at 125 GeV are SM-like, the trilinear coupling can be enhanced by non-vanishing interactions between the detected Higgs boson and heavy BSM scalars. Generally, such couplings can be expressed as:

\begin{equation}
    g_{hh\Phi\Phi} = \frac{2}{v^2}\left(M^2_{\Phi}-\mathcal{M}^2 \right)\,.
\end{equation}

\noindent Here $M_{\Phi}$ is the physical mass of the scalar $\Phi$, $\mathcal{M}$ is a model-specific mass scale controlling the decoupling of the BSM scalars, and $v$ is the SM vacuum expectation value (vev). 
Since the potentially large contributions involving the $g_{hh\Phi\Phi}$-type couplings 
enter the prediction for
$\lambda_{hhh}$ only starting at the loop level, they can give rise to a large effect in comparison to the tree-level coupling, while higher-order loop contributions show the expected perturbative behaviour. 
This situation is 
similar to loop-induced processes such as $h \to \gamma \gamma$ and to the large loop corrections to the mass of the lightest Higgs boson found in supersymmetric models. This implies that the size of the loop corrections stemming from BSM scalars can become as large or even larger than the tree-level prediction, leading to significant BSM deviations in $\kappa_\lambda$.

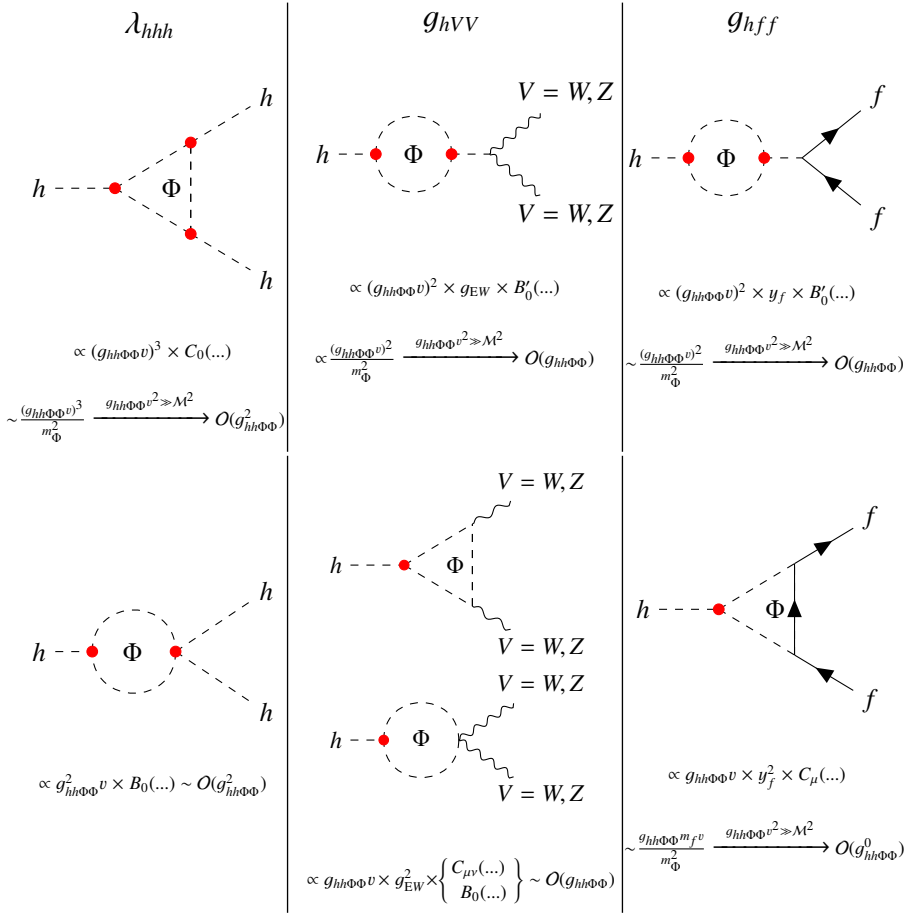
\begin{figure}[t]
\centering
\begin{tikzpicture}
\node (content) {

\begin{minipage}[t]{0.3\textwidth}\centering\vspace{0pt}
\textbf{\large $\lambda_{hhh}$}
\vspace{0.5cm}

\begin{tikzpicture}
\begin{feynman}
    \vertex (h0) at (-1,0){\(h\)};
    \vertex (h1) at (0,0);
    \vertex (h2) at (2,1.2) {\(h\)};
    \vertex (h3) at (2,-1.2) {\(h\)};
    \vertex (h4) at (1,-0.6);
    \vertex (h5) at (1,0.6);

    \diagram* {
      (h0) -- [scalar] (h1),
      (h1) -- [scalar] (h5),
      (h5) -- [scalar] (h2),
      (h1) -- [scalar] (h4),
      (h4) -- [scalar, edge label=\(\Phi\)] (h5),
      (h4) -- [scalar] (h5),
      (h4) -- [scalar] (h3),
    };
    \filldraw[red] ($(h1)!0!(h2)$) circle (2pt) node[below=3pt, red];
    \filldraw[red] ($(h5)$) circle (2pt) node[below=3pt, red];
    \filldraw[red] ($(h4)$) circle (2pt) node[below=3pt, red];
  \end{feynman}
\end{tikzpicture}
\[
{\scriptstyle \propto\;(g_{hh\Phi\Phi }v)^3\; \times\; C_0 (...)}
\]
\[
{\scriptstyle \sim \frac{(g_{hh\Phi\Phi}v)^3}{m_{\Phi}^2} \;\xrightarrow{g_{hh\Phi\Phi}v^2\gg \mathcal{M}^2}\; \mathcal{O}(g_{hh\Phi\Phi}^2)}\;\;
\]
\end{minipage}
\hfill
\begin{minipage}[t]{0.3\textwidth}\centering\vspace{0pt}
\textbf{\large $g_{hVV}$}
\vspace{0.5cm}

\begin{tikzpicture}
\begin{feynman}
    \vertex (h0) at (-1,0){\(h\)};
    \vertex (h1) at (-0.3,0);
    \vertex (h2) at (0.7,0);
    \vertex (h3) at (1.2,0);
    \vertex (v1) at (2.2,0.8) {\(V=W,Z\)};
    \vertex (v2) at (2.2,-0.8) {\(V=W,Z\)};

    \diagram* {
      (h0) -- [scalar] (h1),
      (h1) -- [scalar, half left, looseness=1.7] (h2) -- [scalar, half left, looseness=1.7] (h1),
      (h2) -- [scalar] (h3),
      (h3) -- [boson] (v1),
      (h3) -- [boson] (v2),
    };
    \filldraw[red] ($(h1)!0!(h2)$) circle (2pt) node[below=3pt, red];
    u\filldraw[red] ($(h2)$) circle (2pt) node[below=3pt, red];
    \node at ($(h1)!0.5!(h2) + (0,0)$) {\(\Phi\)};
  \end{feynman}
\end{tikzpicture}
\[
{\scriptstyle \propto\; (g_{hh\Phi\Phi }v)^2\; \times\; g_{\mathrm EW}\; \times\; B'_0 (...)}
\]
\[
{\scriptstyle \propto \frac{(g_{hh\Phi\Phi}v)^2}{m_{\Phi}^2} \;\xrightarrow{g_{hh\Phi\Phi}v^2\gg\mathcal{M}^2}\; \mathcal{O}(g_{hh\Phi\Phi})}
\]
\end{minipage}
\hfill
\begin{minipage}[t]{0.3\textwidth}\centering\vspace{0pt}
\textbf{\large $g_{hff}$}
\vspace{0.5cm}

\begin{tikzpicture}
\begin{feynman}
    \vertex (h0) at (-1,0){\(h\)};
    \vertex (h1) at (-0.3,0);
    \vertex (h2) at (0.7,0);
    \vertex (h3) at (1.2,0);
    \vertex (v1) at (2.2,0.8) {\(f\)};
    \vertex (v2) at (2.2,-0.8) {\(f\)};

    \diagram* {
      (h0) -- [scalar] (h1),
      (h1) -- [scalar, half left, looseness=1.7] (h2) -- [scalar, half left, looseness=1.7] (h1),
      (h2) -- [scalar] (h3),
      (h3) -- [fermion] (v1),
      (v2) -- [fermion] (h3),
    };
    \filldraw[red] ($(h1)!0!(h2)$) circle (2pt) node[below=3pt, red];
    u\filldraw[red] ($(h2)$) circle (2pt) node[below=3pt, red];
    \node at ($(h1)!0.5!(h2) + (0,0)$) {\(\Phi\)};

  \end{feynman}
\end{tikzpicture}
\[
{\scriptstyle \propto\; (g_{hh\Phi\Phi }v)^2\; \times\; y_{f}\; \times\; B'_0 (...)}
\]
\[
{\;\;\scriptstyle \sim \frac{(g_{hh\Phi\Phi}v)^2}{m_{\Phi}^2}\; \xrightarrow{g_{hh\Phi\Phi}v^2\gg\mathcal{M}^2}\; \mathcal{O}(g_{hh\Phi\Phi})}
\]
\end{minipage}
};
\draw[line width=0.5pt]
  ([xshift=-0.17\textwidth]content.north) --
  ([xshift=-0.17\textwidth]content.south);

\draw[line width=0.5pt]
  ([xshift=+0.17\textwidth]content.north) --
  ([xshift=+0.17\textwidth]content.south);

\end{tikzpicture}

\begin{tikzpicture}
\node (content) {
\begin{minipage}{0.3\textwidth}\centering
\begin{tikzpicture}
\begin{feynman}
    \vertex (h0) at (-1,0){\(h\)};
    \vertex (h1) at (-0.3,0);
    \vertex (h3) at (0.8,0);
    \vertex (v1) at (2,0.8) {\(h\)};
    \vertex (v2) at (2,-0.8) {\(h\)};

    \diagram* {
      (h0) -- [scalar] (h1),
      (h1) -- [scalar, half left, looseness=1.7] (h3) -- [scalar, half left, looseness=1.7] (h1),
      (h3) -- [scalar] (v1),
      (h3) -- [scalar] (v2),
    };
    \filldraw[red] ($(h1)$) circle (2pt) node[below=3pt, red];
    u\filldraw[red] ($(h3)$) circle (2pt) node[below=3pt, red];
    \node at ($(h1)!0.53!(h2) + (0,0)$) {\(\Phi\)};

  \end{feynman}
\end{tikzpicture}
\[
{\scriptstyle \propto\;g^2_{hh\Phi\Phi }v\; \times\; B_0 (...)\;\sim\;\mathcal{O}(g_{hh\Phi\Phi}^2)}
\]
\end{minipage}
\hfill
\begin{minipage}{0.3\textwidth}
\centering

\begin{minipage}{0.9\textwidth}
\centering
\begin{tikzpicture}[scale=0.9, transform shape]
\begin{feynman}
    \vertex (h0) at (-1,0){\(h\)};
    \vertex (h1) at (0,0);
    \vertex (h2) at (2,1.2) {\(V=W,Z\)};
    \vertex (h3) at (2,-1.2) {\(V=W,Z\)};
    \vertex (h4) at (1,-0.6);
    \vertex (h5) at (1,0.6);

    \diagram* {
      (h0) -- [scalar] (h1),
      (h1) -- [scalar] (h5),
      (h5) -- [boson] (h2),
      (h1) -- [scalar] (h4),
      (h4) -- [scalar, edge label=\(\Phi\)] (h5),
      (h4) -- [scalar] (h5),
      (h4) -- [boson] (h3),
    };
    \filldraw[red] ($(h1)!0!(h2)$) circle (2pt) node[below=3pt, red];
  \end{feynman}
\end{tikzpicture}
\end{minipage}\hfill%
\begin{minipage}{0.9\textwidth}
\centering
\begin{tikzpicture}[scale=0.9, transform shape]
\begin{feynman}
    \vertex (h0) at (-1,0){\(h\)};
    \vertex (h1) at (-0.3,0);
    \vertex (h3) at (0.8,0);
    \vertex (v1) at (2,0.8) {\(V=W,Z\)};
    \vertex (v2) at (2,-0.8) {\(V=W,Z\)};

    \diagram* {
      (h0) -- [scalar] (h1),
      (h1) -- [scalar, half left, looseness=1.7] (h3) -- [scalar, half left, looseness=1.7] (h1),
      (h3) -- [boson] (v1),
      (h3) -- [boson] (v2),
    };
    \filldraw[red] ($(h1)$) circle (2pt) node[below=3pt, red];
    \node at ($(h1)!0.24!(h2) + (0,-0.25)$) {\(\Phi\)};
\end{feynman}
\end{tikzpicture}
\end{minipage}

\vspace{0.5em}

{\scriptsize
\[
\propto g_{hh\Phi\Phi }v\, \times\; g^2_{\mathrm EW} \times \left\{C_{\mu\nu}(...)\;\atop\;B_0 (...) \right\} \sim 
\mathcal{O}(g_{hh\Phi\Phi})
\]
}

\end{minipage}

\hfill
\begin{minipage}{0.3\textwidth}\centering
\begin{tikzpicture}
\begin{feynman}
    \vertex (h0) at (-1,0){\(h\)};
    \vertex (h1) at (0,0);
    \vertex (h2) at (2,1.2) {\(f\)};
    \vertex (h3) at (2,-1.2) {\(f\)};
    \vertex (h4) at (1,-0.6);
    \vertex (h5) at (1,0.6);

    \diagram* {
      (h0) -- [scalar] (h1),
      (h1) -- [scalar] (h5),
      (h5) -- [fermion] (h2),
      (h1) -- [scalar] (h4),
      (h4) -- [fermion, edge label=\(\Phi\)] (h5),
      (h4) -- [scalar] (h5),
      (h3) -- [fermion] (h4),
    };
    \filldraw[red] ($(h1)!0!(h2)$) circle (2pt) node[below=3pt, red];
  \end{feynman}
\end{tikzpicture}
\[
{\scriptstyle \propto\; g_{hh\Phi\Phi }v\; \times\; y_{f}^2\; \times\; C_\mu (...)}
\]
\[
{\;\;\scriptstyle \sim \frac{g_{hh\Phi\Phi}m_fv}{m_{\Phi}^2}\; \xrightarrow{g_{hh\Phi\Phi}v^2\gg\mathcal{M}^2}\; \mathcal{O}(g^0_{hh\Phi\Phi})}
\]
\end{minipage}
};
\draw[line width=0.5pt]
  ([xshift=-0.17\textwidth]content.north) --
  ([xshift=-0.17\textwidth]content.south);

\draw[line width=0.5pt]
  ([xshift=+0.17\textwidth]content.north) --
  ([xshift=+0.17\textwidth]content.south);

\end{tikzpicture}
\caption{Powers of the large coupling $g_{hh\Phi\Phi}$ (marked in red) contributing at the one-loop order to the couplings involving different numbers of SM-like Higgs bosons. In columns from left to right: the trilinear Higgs coupling, $\lambda_{hhh}$, the coupling of a Higgs boson to two gauge bosons, $g_{hVV}$, and the couplings of a Higgs boson to two fermions $g_{hff}$. Below each diagram we indicate the proportionality to the couplings of interest, the loop functions and the order of contribution of $g_{hh\Phi\Phi}$ in the limit where mass splitting effects are large, namely where $g_{hh\Phi\Phi}v^2 \gg \mathcal{M}^2$ (which implies $m_\Phi^2\sim g_{hh\Phi\Phi}v^2/2$). 
}
\label{fig:pcounting}
\end{figure}

The presence of large radiative corrections to the trilinear Higgs self-coupling naturally prompts two questions: 1) how large can higher-order effects can become in other Higgs observables that depend on the $g_{hh\Phi\Phi}$ interaction, and 2) which observable would be expected to exhibit the first measurable deviation from the SM in such scenarios? Simple power-counting arguments, illustrated in Fig.~\ref{fig:pcounting}, indicate that, in the regime of large $ g_{hh\Phi\Phi}$, the leading one-loop BSM contribution to $\lambda_{hhh}$ scales as $\mathcal{O}( g^2_{hh\Phi\Phi})$, whereas corrections to single-Higgs couplings grow at most linearly with $ g_{hh\Phi\Phi}$.

Complementary SMEFT-based considerations lead to the same qualitative hierarchy between deviations in trilinear and single-Higgs interactions~\cite{Durieux:2022hbu}.

In the following we discuss specific models as examples of the above-mentioned phenomenology. We denote with $h$ the Higgs boson discovered at 125 GeV, while other BSM scalars are assumed to be heavier. In all the scenarios considered, we have checked that the displayed parameter space is compatible with state-of-the-art theoretical constraints such as perturbative unitarity and vacuum stability, as well as with current experimental constraints --- in particular those arising from electroweak precision data, direct collider searches, and Higgs signal-strength measurements.


\section{Results}\label{sec2}
As a first example, we discuss the $\mathbf{Z}_2$-symmetric singlet extension of the SM ($\mathbf{Z}_2$SSM).\footnote{We refer the reader to, e.g., Ref.~\cite{Braathen:2020vwo} for an overview of the notations employed in this work. We specifically consider here the $N=1$ case of the $O(N)$-symmetric SSM of Ref.~\cite{Braathen:2020vwo}.} The symmetry in the $\mathbf{Z}_2$SSM prevents the BSM scalar, denoted $S$, from mixing with $h$. 
Its mass is given by $m_S^2=\mu_S^2+\lambda_{S\Phi}v^2$, where $\mu_S$ is the singlet mass parameter in the Lagrangian and $\lambda_{S\Phi}$ is the portal coupling between the singlet and the SM-like doublet, which plays the role of the large $g_{hh\Phi\Phi}$ coupling introduced above. At the one-loop level the couplings of the SM-like Higgs with vectors (and fermions), e.g. $g_{hZZ}$, receive only external-leg corrections --- vertex corrections are forbidden by the unbroken $\mathbf{Z}_2$ symmetry. Since there are no mixing effects 
nor any vertex-type corrections, 
the one- and two-loop expressions for both $g_{hZZ}$ and $\kappa_{\lambda}$ are particularly compact~\cite{BahlhZZ}.  At two loops, the correction to $g_{hZZ}$ is not entirely due to the external-leg contribution. An additional effect arises from the on-shell (OS) renormalization of the electroweak vev, which enters when expressing single-Higgs couplings as masses divided by the vev. The on-shell (OS) vev renormalization involves the $W$ and $Z$ self-energies. While these do not receive a contribution from $S$ at one-loop --- since $S$ is a singlet and does not directly couple to the electroweak gauge bosons --- a two-loop contribution is present. This term is essential to cancel the renormalisation-scale dependence of the external-leg correction.

\begin{figure}[h]
\centering
\includegraphics[width=0.48\textwidth]{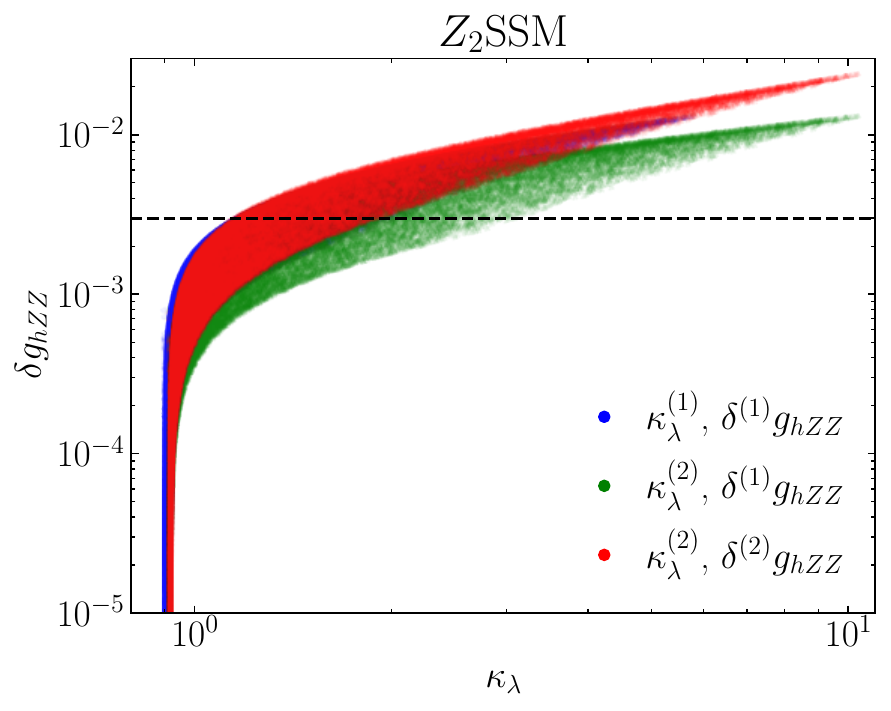}
\includegraphics[width=0.51\textwidth]{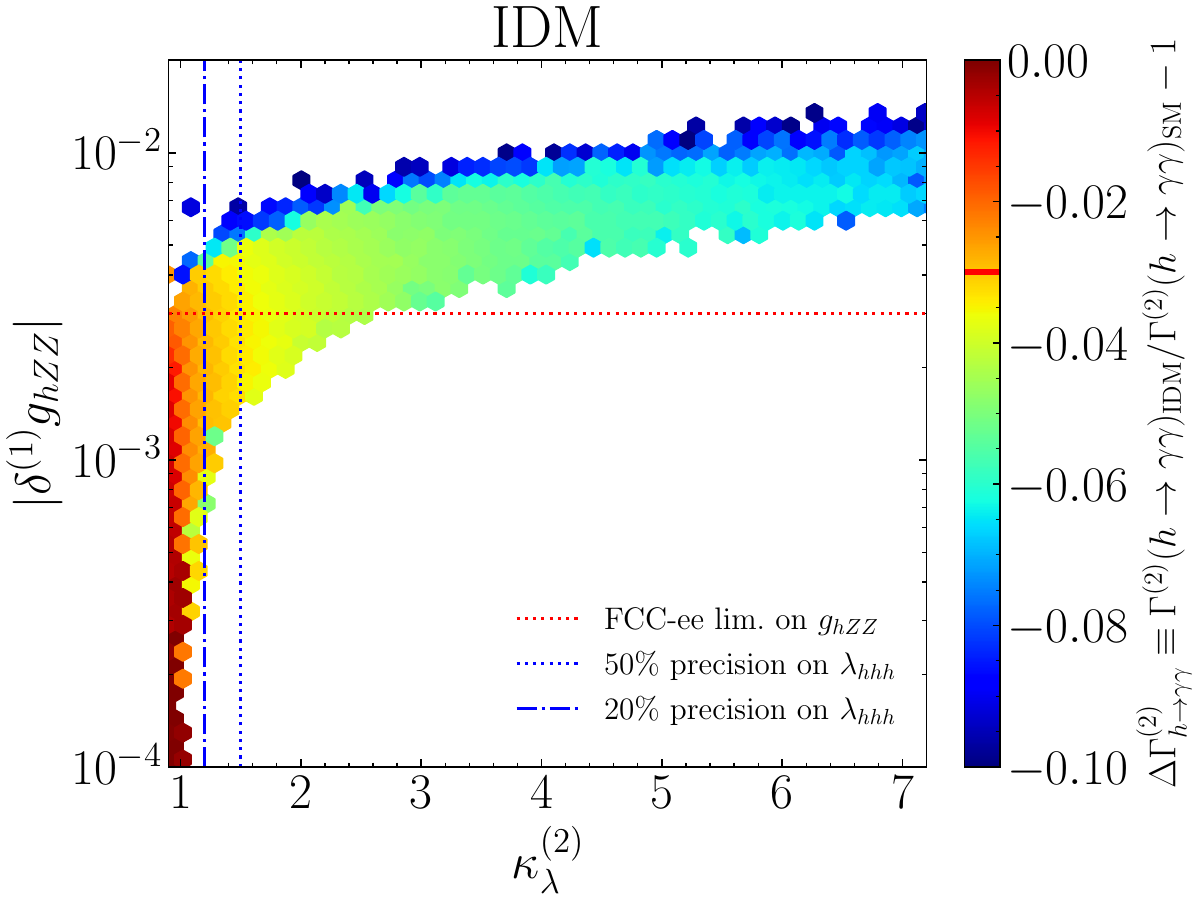}
\caption{\textit{Left}: Parameter scan results for the $\mathbf{Z}_2$SSM in the plane of $\kappa_\lambda$ and $\delta g_{hZZ}$. The colour of the points indicates the perturbative order (one or two loops) at which the two quantities are computed, as specified in the legend. The black dashed line shows the expected $2\sigma$ sensitivity of FCC-ee to the $g_{hZZ}$ coupling. \textit{Right}: Parameter scan results in the IDM in the plane of $\kappa_\lambda^{(2)}$ (at two loops) and $\delta^{(1)}g_{hZZ}$ (at one loop). Points are grouped into hexagonal bins, whose colour indicates the minimal BSM deviation in $\Gamma(h\to\gamma\gamma)$ among the points in each bin. The red solid line in the colour bar corresponds to the expected $2\sigma$ FCC-ee sensitivity to $h\to\gamma\gamma$~\cite{Selvaggi:2025kmd}. Additional reference lines are described in the legend and in the text. }
\label{zsm_idm}
\end{figure}

In the left panel of \reffi{zsm_idm} we show the results of a parameter scan of the allowed region of the $\mathbf{Z}_2$SSM parameter space. On the horizontal axis 
$\kappa_{\lambda}$ is displayed, while the vertical axis shows the BSM deviation in the $g_{hZZ}$ coupling, denoted $\delta g_{hZZ}$. The perturbative order of the calculation is indicated in the legend via a superscript: blue points correspond to one-loop predictions for both $\kappa_{\lambda}^{(1)}$ and $\delta^{(1)} g_{hZZ}$; green points show $\kappa_{\lambda}^{(2)}$ at two loops and $\delta^{(1)} g_{hZZ}$ at one loop; and red points display both quantities at two loops, $\kappa_{\lambda}^{(2)}$ and $\delta^{(2)} g_{hZZ}$. Since the red points are plotted on top of the others, they cover most of the blue region and partially the green one. We observe that large shifts in $\kappa_{\lambda}$ are possible --- for example, up to $\kappa_{\lambda}^{(2)}\approx 3\ (2)$, if one-loop (two-loop) corrections to $g_{hZZ}$ are taken into account --- in parameter regions where the corresponding deviation in $g_{hZZ}$ remains below the $2\sigma$ sensitivity at FCC-ee (derived from \citere{Selvaggi:2025kmd}), indicated by the horizontal black dashed line. We also recall that the one- and two-loop predictions for $\kappa_\lambda$ in the SM (including for simplicity only the dominant contributions involving the top quark, taking into account at the two-loop level the QCD and leading electroweak contributions) are $\kappa_\lambda^{\text{SM, }(1)}\simeq 0.91$ and $\kappa_\lambda^{\text{SM, }(2)}\simeq 0.92$~\cite{Braathen:2019zoh}.

As a second example, we consider the Inert Doublet Model (IDM), a $\mathbf{Z}_2$-symmetric variant of the 2HDM in which the second scalar doublet does not acquire a vev and does not mix with the SM-like Higgs boson. We refer the reader to Refs.~\cite{Deshpande:1977rw,Barbieri:2006dq,Braathen:2019zoh,Aiko:2023nqj} for comprehensive reviews of the model. Similarly to the $\mathbf{Z}_2$SSM, the masses of the BSM scalars $\Phi=H,\ A,\ H^\pm$ of the IDM can be written as $m_\Phi^2=\mu_2^2 +\frac{1}{2}\lambda_\Phi v^2$, where $\lambda_\Phi$ denotes a combination of the Lagrangian quartic couplings and $\mu_2$ sets the BSM mass scale. In contrast to the $\mathbf{Z}_2$SSM, single-Higgs couplings in the IDM do receive genuine BSM vertex corrections, which we have evaluated consistently at the one-loop level (
augmented by leading two-loop corrections from Ref.~\cite{Aiko:2023nqj} included for the partial decay width of the Higgs to two photons).

The scan results are shown in the right panel of~\reffi{zsm_idm}. The horizontal axis displays the two-loop prediction for $\kappa_\lambda$ (from Refs.~\cite{Braathen:2019pxr,Braathen:2019zoh,Aiko:2023nqj}), while the vertical axis shows the one-loop BSM deviation in the $g_{hZZ}$ coupling. For the latter, both BSM vertex and external-leg corrections are included, evaluated at vanishing external momenta in order to obtain a quantity corresponding to $\kappa_Z$~\cite{LHCHiggsCrossSectionWorkingGroup:2012nn, LHCHiggsCrossSectionWorkingGroup:2013rie}. The colour of each hexagonal bin represents the minimal BSM deviation in the Higgs partial decay width to two photons, $\Gamma(h\to\gamma\gamma)$, at two loops among the points within the bin. This additional observable is relevant for the IDM due to the presence of a charged BSM Higgs boson in the model. The vertical blue lines correspond to $20\%$ (dot-dashed) and $50\%$ (dotted) levels of accuracy on the determination of $\lambda_{hhh}$~(taking $\kappa_\lambda = 1$ as central value), while the red dotted horizontal corresponds to the $2\sigma$ precision expected for $g_{hZZ}$ at FCC-ee. We observe that a significant population of our scan points exhibit significant BSM deviations in $\kappa_\lambda$, reaching values up to $\kappa_\lambda\sim 2.8$, while remaining below the $2\sigma$ FCC-ee sensitivity on $g_{hZZ}$. For most of these points, a measurement of $\Gamma(h\to\gamma\gamma)$ at the FCC-ee would also remain compatible with the SM up to the $1.5\sigma$ level.

Finally, we have investigated the impact of including momentum dependent effects as well as diagrams involving an insertion of the one-loop modified value of $\kappa_\lambda$ in the calculation of the $g_{hZZ}$ coupling. While these contributions are formally of two-loop order, their size becomes relevant for large $\kappa_\lambda$. For the IDM parameter points considered here, we find that both effects tend to reduce the overall BSM deviation in $g_{hZZ}$~\cite{BahlhZZ}. The results shown in \reffi{zsm_idm} therefore provide a conservative estimate of the possible deviations in $\kappa_\lambda$ possible for IDM scenarios with BSM effects in $g_{hZZ}$ below the $2\sigma$ sensitivity of the FCC-ee.

\medskip
As a third example, we explore the general real singlet scalar extension of the SM (RxSM), previously studied in~\citeres{Barger:2007im,Lerner:2009xg,Costa:2015llh,Li:2019tfd,Arco:2025nii,Braathen:2025qxf,Braathen:2025svl}. 
In this context it is interesting to note that the model served as the exemplary case for the relation between $\kala$ and $g_{hZZ}$ 
in the recent report of the ``European Strategy for Particle Physics Update'' (ESPPU)~\cite{deBlas:2025gyz}. 
In our analysis we go beyond some of the approximations made in that reference, and thus our results on the RxSM can yield additional 
information not taken into account in \citere{deBlas:2025gyz}.


\begin{figure}[h]
\centering
\includegraphics[width=0.49\textwidth]{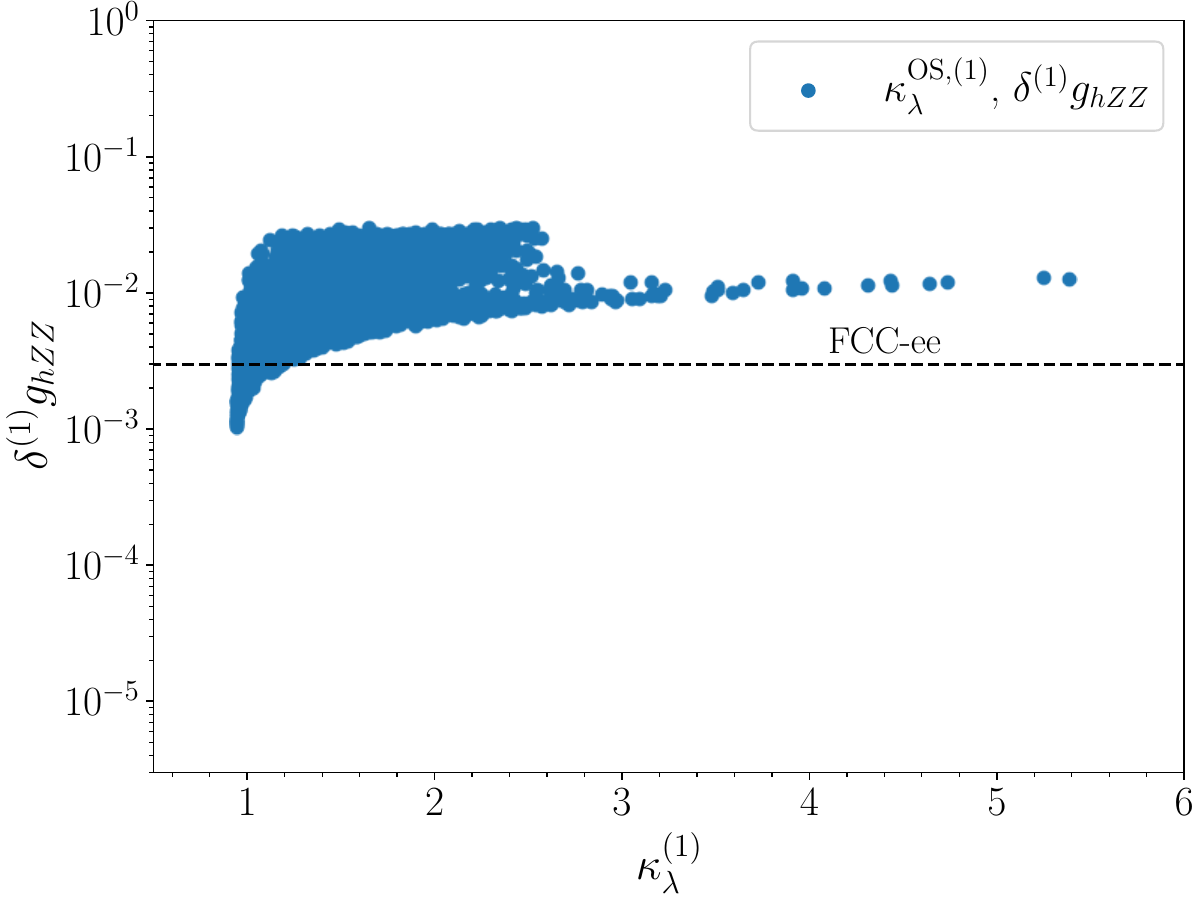}
\caption{
One-loop OS results for the $g_{hZZ}$ coupling in the RxSM, normalised to the SM one-loop value, as a function of the one-loop OS prediction for $\kappa_\lambda$.}
\label{rxsm}
\end{figure}

In Fig.~\ref{rxsm} we present results from a general parameter scan of the general (i.e.\ non-$\mathbb{Z}_2$ symmetric) singlet extension of the SM, the RxSM, where all current theoretical and experimental constraints are imposed, in particular \texttt{HiggsBounds} and \texttt{HiggsSignals}~\cite{Bahl:2022igd} constraints, as well as constraints from electroweak precision observables ($S$ and $T$~\cite{Peskin:1991sw}).
The black dashed line indicates the projected $2\sigma$ exclusion reach of 
FCC-ee. 
The results shown here employ an on-shell (OS) renormalisation scheme for $\delta^{(1)}g_{hZZ}$. The $g_{hZZ}$ coupling is normalized to the SM one-loop prediction, in order to isolate genuine BSM effects and avoid including SM radiative corrections into the definition of the coupling modifier. The OS definition of $g_{hZZ}$ is obtained by supplementing the $\overline{\mathrm{MS}}$ result~\footnote{We have found that in the $\overline{\mathrm{MS}}$-scheme there is a strong dependence on the renormalisation scale that was not considered in Ref.~\cite{Huang:2016cjm}.  
In the displayed plot we
adopted an OS scheme 
for which the result is not affected by a renormalisation 
scale dependence.} with OS counterterms for the electroweak vev and the scalar mixing angle, following the same renormalisation prescription as adopted for $\kappa_\lambda$ in Ref.~\cite{Braathen:2025qxf}. 
In this scenario, deviations of $\kappa_{\lambda}$ of up to $40\%$ remain compatible with deviations in $g_{hZZ}$ below the FCC-ee sensitivity. 
Because in the specific scenario of the RxSM the loop effects on different quantities are largely correlated to each other, we find in this case that
the size of the corrections to $\kappa_\lambda$ that are possible 
in the parameter region where
the measurement of $g_{hZZ}$ at FCC-ee is compatible with the SM is smaller than in the other models we have considered (see also the 2HDM case below). This indicates that drawing conclusions on the relative size of corrections to Higgs couplings from the RxSM alone (as done in Ref.~\cite{deBlas:2025gyz}) 
is not sufficient for representing the possible phenomenology of extended Higgs sectors. 

\begin{figure}[h]
\centering
\includegraphics[width=0.49\textwidth]{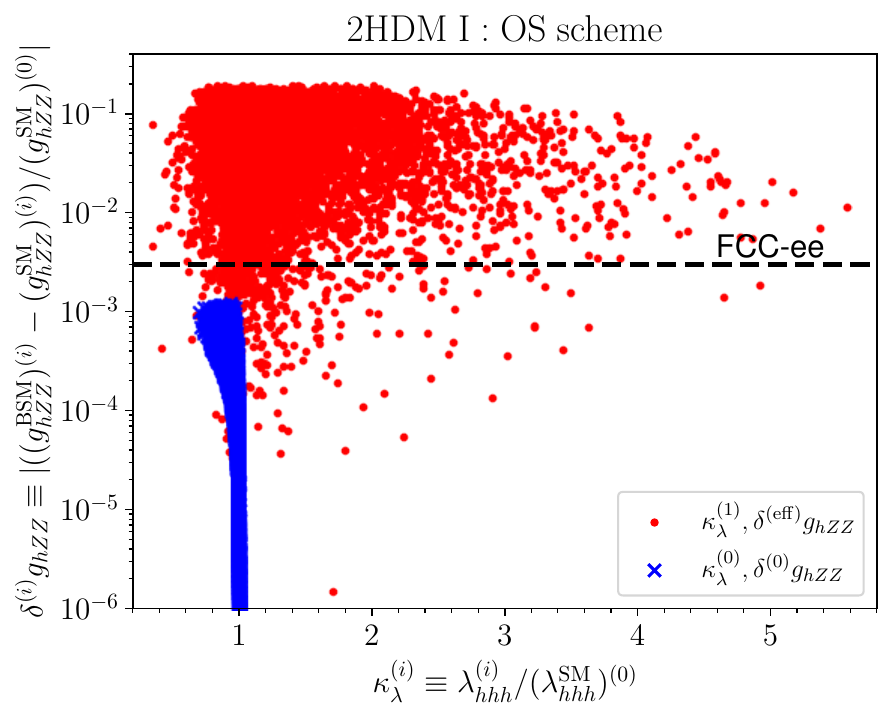}
\includegraphics[width=0.49\textwidth]{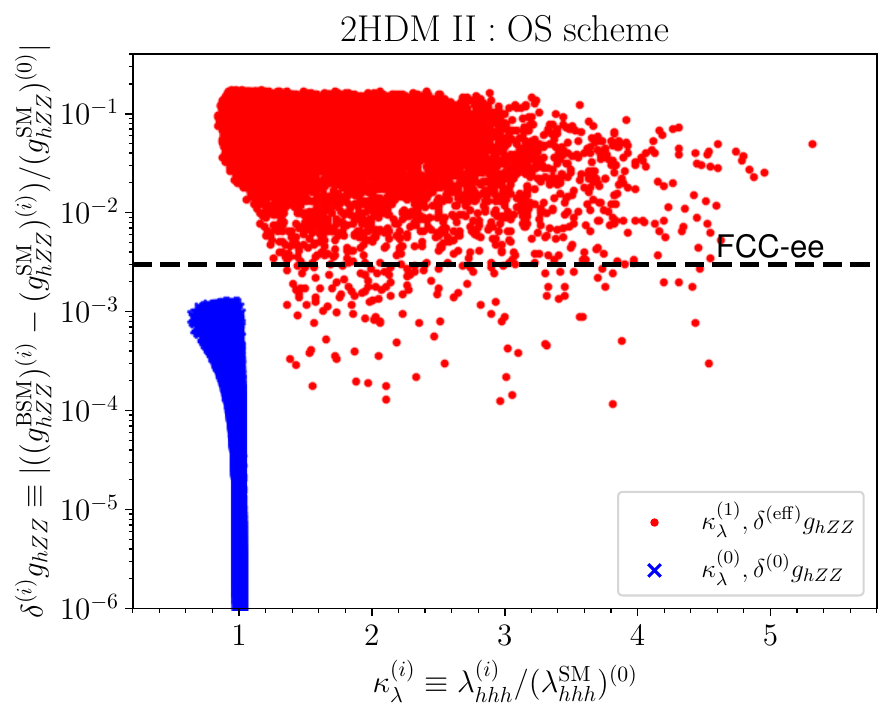}
\caption{Parameter scan results in the plane \{$\kala$, $g_{hZZ}$\} in the 2HDM-I (left) and 2HDM-II (right). Blue points indicate the tree-level results for these couplings, while red points show for the same parameter points the one-loop predictions including also for $g_{hZZ}$ the leading higher-order corrections 
involving insertions of one or two powers of the trilinear coupling, see text for details. The black dashed lines indicate the projected sensitivity at 
FCC-ee.}\label{thdmeff}
\end{figure}

\medskip
Finally, we analyse the 2HDM; for reviews see~\citeres{Gunion:1989we,TDLee,Aoki:2009ha,Branco:2011iw}. \reffi{thdmeff} shows the allowed parameter space for Yukawa Types I (left) and II (right). Blue points correspond to tree-level predictions for $\kappa_{\lambda}$ and the $g_{hZZ}$ coupling, while red points show the corresponding one-loop values for the same points. 

The trilinear coupling is computed with \texttt{anyH3}~\cite{Bahl:2023eau} using the KOSY scheme~\cite{Kanemura:2004mg} (implemented in the develop branch~\cite{anyHH}), while the predictions for the $g_{hZZ}$ coupling include the full one-loop corrections evaluated in an OS scheme with the mixing-angle counterterms defined according to the the KOSY scheme. Leading higher-order effects are included via an effective one-loop value of $\kappa_\lambda$, involving in particular vertex-like corrections with one power of $\kappa^{(1)}_{\lambda}$ and external-leg corrections with two powers of $\kappa^{(1)}_{\lambda}$.\footnote{The result including this formally higher-order corrections is UV finite and captures the leading corrections.} These corrections significantly enlarge the allowed region in the $g_{hZZ}$–$\kappa_\lambda$ plane, with $\kappa_\lambda$ values spanning $0.4 <\kappa_{\lambda}<5.8$ in Type I and $0.8 <\kappa_{\lambda}<5.5$ in Type II. While scan dependent, these ranges are consistent with the results of Ref.~\cite{Arco:2025pgx}, which were computed using the public tool \texttt{BSMPT}~\cite{Basler:2018cwe,Basler:2020nrq,Basler:2024aaf}. On the other hand, the effective $g_{hZZ}$ coupling can deviate by up to $20\%$ in both types, which is nearly two orders of magnitude larger than the largest deviations found at tree level. The presence of parameter points below the FCC-ee sensitivity curve for values of $\kappa_{\lambda}$ larger than the projected HL-LHC $2\sigma$ sensitivity (at $\kappa_{\lambda} \sim 1.6$~\cite{ATLAS:2025eii}) demonstrates that new-physics effects can manifest themselves first in the trilinear Higgs self-coupling. Moreover, we also find that a significant population of points occurs both below the sensitivity of FCC-ee on $g_{hZZ}$ and of HL-LHC on $\kala$; a large fraction of these points could however be probed via a high-precision measurement of the trilinear self-coupling. 

As an additional theoretical motivation for the 2HDM, we consider the possibility of realizing a strong first-order electroweak phase transition. The relevant parameter points are obtained using the private code of Ref.~\cite{Biekotter:2022kgf}, based on a four-dimensional effective potential including one-loop Coleman–Weinberg corrections, thermal effects, and daisy resummation in the Arnold–Espinosa scheme~\cite{Arnold:1992rz}.

\begin{figure}[h]
\centering
\includegraphics[width=0.49\textwidth]{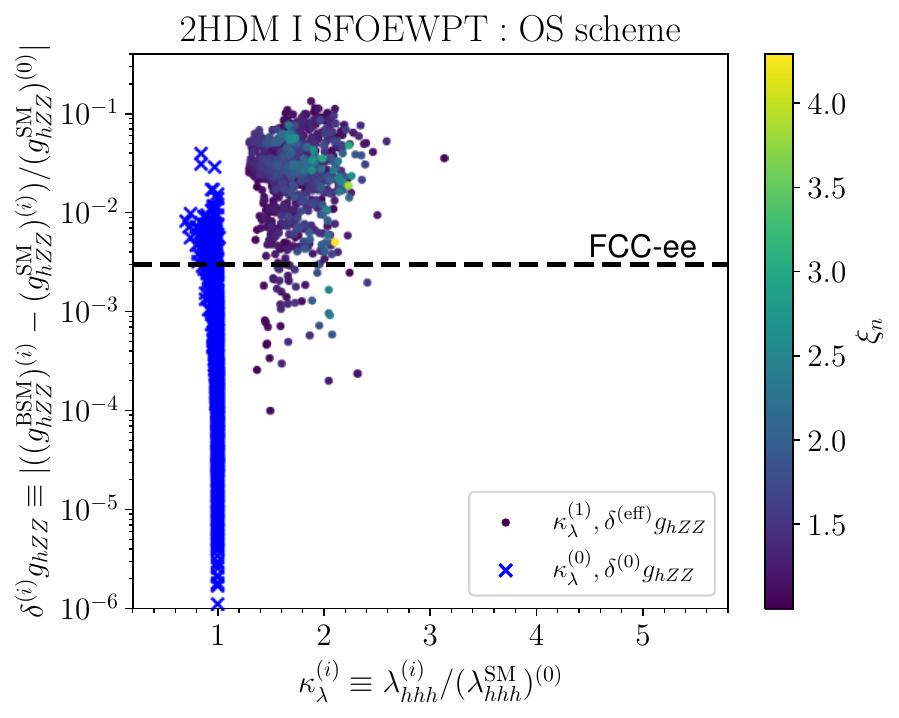}
\includegraphics[width=0.49\textwidth]{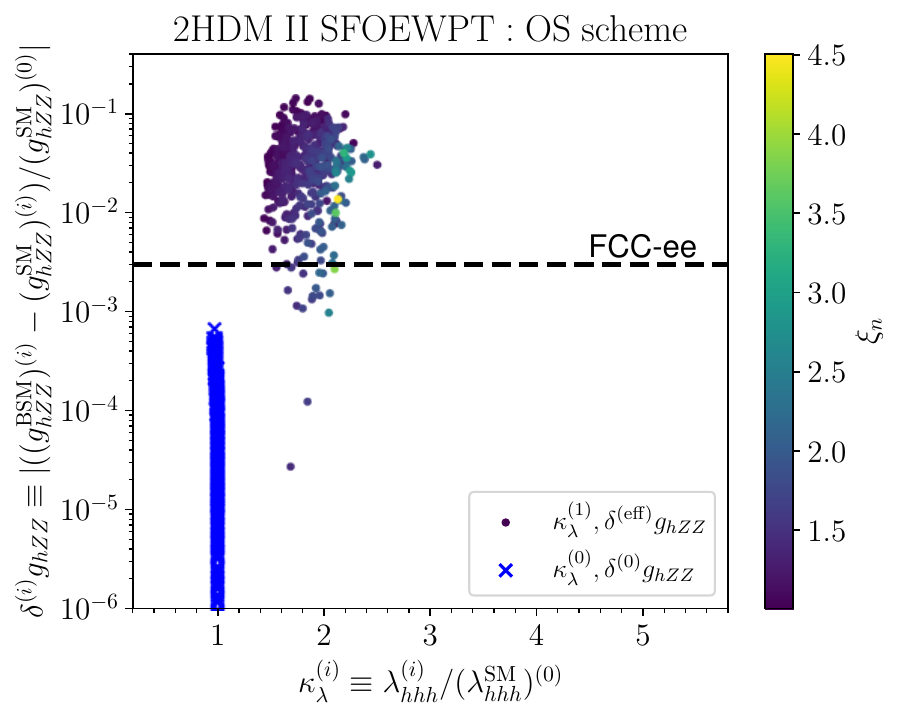}
\caption{Same as Fig.~\ref{thdmeff}, but restricted to parameter points featuring a SFOEWPT. The colour coding indicates the strength of the transition $\xi_n$.}\label{thdmsfoewpt}
\end{figure}

Imposing the SFOEWPT criterion restricts the allowed range of $\kappa_\lambda$ without significantly modifying the allowed values of $g_{hZZ}$, as shown in Fig.~\ref{thdmsfoewpt}. The colour coding indicates the strength of the phase transition, $\xi_n$, with successful baryogenesis requiring $\xi_n \gtrsim 1$.  We observe that the strongest transitions are realised for the largest values of $\kappa_{\lambda}$, as those are correlated to a larger barrier between the false and the electroweak minima. Our results are consistent with Fig.~11 of Ref.~\cite{Biekotter:2022kgf}, where smaller values of $\kappa_\lambda$ fail to produce a sufficiently strong transition, while larger values are excluded due to vacuum trapping. Once again, the existence of points below the FCC-ee sensitivity in $g_{hZZ}$ but above the HL-LHC reach in $\kappa_\lambda$ highlights the importance of loop effects in the Higgs self-coupling.%
\footnote{
Our conclusions differ from \citere{Anisha:2025zbc}, where all scanned points were found to yield deviations in $g_{hZZ}$ above the FCC-ee
sensitivity. This can be attributed to different renormalisation schemes employed in the two calculations, and to the inclusion of the projected theory
uncertainties to EWPOs in our analysis. Furthermore, in \citere{Anisha:2025zbc} the $e^+e^- \to Zh$ cross section is evaluated, whereas in our analysis we 
restrict ourselves to the higher-order corrections to $g_{hZZ}$. However, we consider this difference to have a minor effect, and we trace the discrepancy
in our conclusions to the 
differences in the EWPO constraints. 
}
Finally, we recall that if the value of $\kala$ that is realised in nature is $\sim 2$, the absolute accuracy of the bounds from HL-LHC would be noticeably degraded compared to the case of $\kala\sim 1$~\cite{ATLAS:2025eii}, due to the decrease of the di-Higgs production cross-section. This implies that for these 2HDM points with a SFOEWPT only a moderate deviation from the SM predictions 
could be seen at the HL-LHC. Moreover, while a SFOEWPT in the 2HDM is always associated with a significant BSM deviation in $\kala$, this is not the case for all models that can accommodate a SFOEWPT, especially if the first-order phase transition does not occur along the SM-like doublet field direction (e.g.\ in the RxSM for singlet-driven SFOEWPT~\cite{Braathen:2025svl} or in the cxSM~\cite{Biekotter:2025npc}, see also \citeres{Ramsey-Musolf:2019lsf,Niemi:2020hto} for the case of a triplet model). 

\section{Conclusions}\label{conclusions}

In our work we have shown that there exist scenarios within typical BSM frameworks in which deviations in single-Higgs couplings (such as in particular $g_{hZZ}$) are too small to be observed at future $e^+e^-$ colliders, while deviations in the trilinear Higgs self-coupling $\kappa_\lambda$ can be sizeable and potentially detectable already at the HL-LHC or at an $e^+e^-$ Higgs factory. Our results highlight the importance of direct and model-independent measurements of the
trilinear Higgs coupling. 
A next-generation Higgs factory should therefore aim for the highest possible precision in probing this key parameter, for any central 
value of $\kappa_\lambda$ that can be realised (taking into account the theoretical and experimental constraints) in BSM models.

\section*{Acknowledgements}
The work of S.H.\ has received financial support from the
grant PID2019-110058GB-C21 funded by
MCIN/AEI/10.13039/501100011033 and by ``ERDF A way of making Europe'', as well as from Grant PID2022-142545NB-C21 funded by MCIN/AEI/10.13039/501100011033/ FEDER, UE.
S.H.\ and K.R. acknowledge the support of the Spanish Agencia Estatal de Investigación through the grant “IFT Centro de Excelencia Severo
Ochoa CEX2020-001007-S”. The project that gave rise to these results received the support of a fellowship from the “la Caixa” Foundation (ID 100010434). The fellowship code is
LCF/BQ/PI24/12040018. J.B., A.V.S.\ and G.W.\ acknowledge support by the Deutsche Forschungsgemeinschaft (DFG, German Research Foundation) under Germany's Excellence Strategy --- EXC 2121 ``Quantum Universe'' --- 390833306. This work has been partially funded by the Deutsche Forschungsgemeinschaft (DFG, German Research Foundation) --- 491245950. J.B. and A.V.S are supported by the DFG Emmy Noether Grant No.\ BR 6995/1-1.

\bibliography{bibliography.bib}

\end{document}